%% file: main.tex
\documentclass[pdflatex,sn-mathphys-num, twocolumn, ]{sn-jnl}
\input{tex/packages}

\title[Quantum Simulation of Magnetic Materials: from Ab-Initio to NISQ]{Quantum Simulation of Magnetic Materials: from Ab-Initio to NISQ}
\author[1]{Pascal Stadler~\orcidlink{0000-0001-9873-0041}}
\author[1]{Florian G. Eich~\orcidlink{0000-0002-0434-6100}}
\author[1]{Benedikt M. Schoenauer~\orcidlink{0000-0002-6089-4398}}
\author[1]{Peter Schmitteckert~\orcidlink{0000-0001-9642-5115}}
\author[1]{Michael Marthaler}
\author[2]{Gary Schmiedinghoff~\orcidlink{0000-0003-2259-7365}}
\author[2]{Peter K. Schuhmacher~\orcidlink{0000-0003-1232-4363}}
\author[1]{Sebastian Zanker~\orcidlink{0000-0002-9097-740X}}
\affil[1]{\orgname{HQS Quantum Simulations GmbH}, \orgaddress{\street{Rintheimer Straße 23}, \postcode{76131} \city{Karlsruhe}, \state{Germany}}}
\affil[2]{\orgdiv{Department of High Performance Computing, Institute of Software Technology}, \orgname{German Aerospace Center~(DLR)}, \orgaddress{\street{Linder Höhe}, \postcode{51147} \city{Cologne}, \state{Germany}}}

\begin{document}

\abstract{
Quantum computers are increasingly accessible, yet demonstrations of physically meaningful simulations for real materials remain scarce. In our work we simulate low-energy magnetic excitations, specifically spin-wave spectra, of chromium tri-halide monolayers. Starting from ab-initio electronic structure calculations for these two-dimensional magnets, we derive an effective spin model and simulate low-energy spin excitations using a real-time propagation of the spin system on the commercial quantum computing cloud platform IQM Resonance. The results for systems with up to 48 qubits are validated against classical benchmarks.
While some spectral features remain challenging for today's NISQ devices, our simulation achieves good agreement at quasi-constant wall-time scaling, compared to the exponential scaling of classical methods.
Our results demonstrate that, even in the absence of quantum advantage, useful quantum simulations of real materials are becoming possible for domain experts via commercial cloud access to quantum computers.
}

\keywords{Quantum simulation of materials, Cloud-accessed quantum computing, Ab-Initio quantum workflow, Quantum utility, Two-dimensional magnets, Spin-wave dynamics}

\maketitle

\input{content/Motivation}
\input{content/Workflow}
\input{content/Results}
\input{content/Conclusion}
\input{content/Methods}

\section*{Acknowledgements}
\input{adds/acknowledgements}

\section*{Data availability}
\input{adds/data_availability}

\section*{Code availability}
\input{adds/code_availability}

\section*{Author contributions}
\input{adds/author_contributions}

\section*{Competing interests}
\input{adds/competing_interests}

\bibliography{references}

\end{document}

%% file: tex/packages.tex
\usepackage{hyperref}
\usepackage{tikz}

\usepackage{graphicx}%
\usepackage{multirow}%
\usepackage{amsmath,amssymb,amsfonts}%
\usepackage{amsthm}%
\usepackage{mathrsfs}%
\usepackage[title]{appendix}%
\usepackage{xcolor}%
\usepackage{textcomp}%
\usepackage{manyfoot}%
\usepackage{booktabs}%
\usepackage{algorithm}%
\usepackage{algorithmicx}%
\usepackage{algpseudocode}%
\usepackage{listings}
\usepackage[capitalise]{cleveref}

\usepackage{orcidlink}

%% file: content/Motivation.tex
\section{Motivation}
Quantum computing is on the cusp of transitioning from theoretical promise to practical utility, 
driven by unprecedented advances in cloud-accessible quantum processors.
Recent demonstrations \cite{GoogleQ2025, Zhang2025_nmr_otoc_geometry, Roncevic2026_half_mobius, Alam2025_fermionic_dynamics, Maskara2025_reconfigurable_materials, Lee2026_neutron_scattering_benchmark} highlight a pivotal shift: quantum computers are becoming useful, even without quantum advantage -- sometimes referred to as quantum utility~\cite{Kim2023Utility}. 
The idea is to focus on problems in physics that are hard to solve using brute-force approaches on conventional computers and can be addressed using quantum computers without requiring a proven quantum advantage. 
This promotes quantum computers as additional generic tool for simulating quantum physics, in contrast to classical algorithms specialized for a certain class of problems.

As hardware capabilities advance, the focus has shifted toward identifying real-world applications, 
both to demonstrate quantum utility and to validate the entire hardware-software stack in a context relevant in physics beyond the studies of quantum computation itself \cite{Zimboras2025}.
In particular, domain experts require algorithms that can provide useful data when accessing quantum hardware via the cloud, without on-site access and expert knowledge to manually tune the quantum hardware to their use case.
In particular, end-to-end workflows connecting first-principles modeling and quantum simulation are an important direction for practical quantum computing applications.

The most compelling candidate for early quantum utility 
in a realistic and industrially relevant setting 
is the simulation of quantum systems~\cite{Tazhigulov2022_sycamore_materials, Stanisic2022_fermi_hubbard, RobledoMoreno2025_quantum_centric_chemistry,
Selisko2025_dmft_real_materials,
Kandala2017_vqe_quantum_magnets,Arute2020_charge_spin_fermi_hubbard, Gajewski2025}.
Two-dimensional magnetic materials, which only during the last decade have been confirmed experimentally \cite{Gong2017_2d_magnet_discovery, Huang2017_2d_magnet_discovery}, provide an interesting test bed for such quantum simulations, since they naturally map to qubits due to their magnetic nature. Furthermore, the reduced dimensionality implies a low qubit connectivity, enabling the quantum simulation on Noisy Intermediate-Scale Quantum (NISQ) devices. 
These devices can compute fundamental excitations of these two-dimensional magnets, such as magnons or spin waves, i.e., collective precessions of the spins, by simulating the time propagation after an initial quench.
Crucially, magnon dispersion relations can be directly probed 
via inelastic neutron scattering \cite{Chen2018_CrI3_topological_neutron_scattering, Cai2021_CrBr3_topological_neutron_scattering, Chen2022_CrCl3_topological_neutron_scattering} and other experimental techniques,\cite{LeTacon2011_RIXS_paramagnon,Yoshihara2023_BLS_magnetic_excitations,Kepaptsoglou2025_STEM_EELS_magnons}
enabling rigorous benchmarking of quantum simulations 
beyond classical simulation capabilities. 

In this work, we present a first-principles-driven workflow 
for simulating low-energy spin-wave spectra in chromium tri-halide monolayers. 
Starting from the structural information of the material, we derive an effective spin model using parity optimization \cite{schoenauer25_spin_mapper} to identify spin-like orbitals, project to a low-energy subspace, and encode magnon wave vectors via ``twisted'' spin-spin interactions. 
The resulting spin chain is simulated through IQM Resonance~\cite{IQMResonance} on Garnet and Emerald chips at high-symmetry wave vectors, labeled $\Gamma$, K and M, of the magnetic excitation by performing a time evolution after an initial quench for up to 48 spins.
We benchmark the resulting spin-wave spectra against classical, Krylov-based simulations for system sizes up to 24 spins and density matrix renormalization group (DMRG) calculations for system sizes up to 48 spins and demonstrate the quasi-constant wall-time scaling of our simulation.
Our quantum simulations are performed via the cloud, without direct access to the hardware, proving that quantum utility for material science is already within reach of researchers. 

%% file: content/Workflow.tex
\begin{figure*}[!t]
\centering
\includegraphics[width=0.95\linewidth]{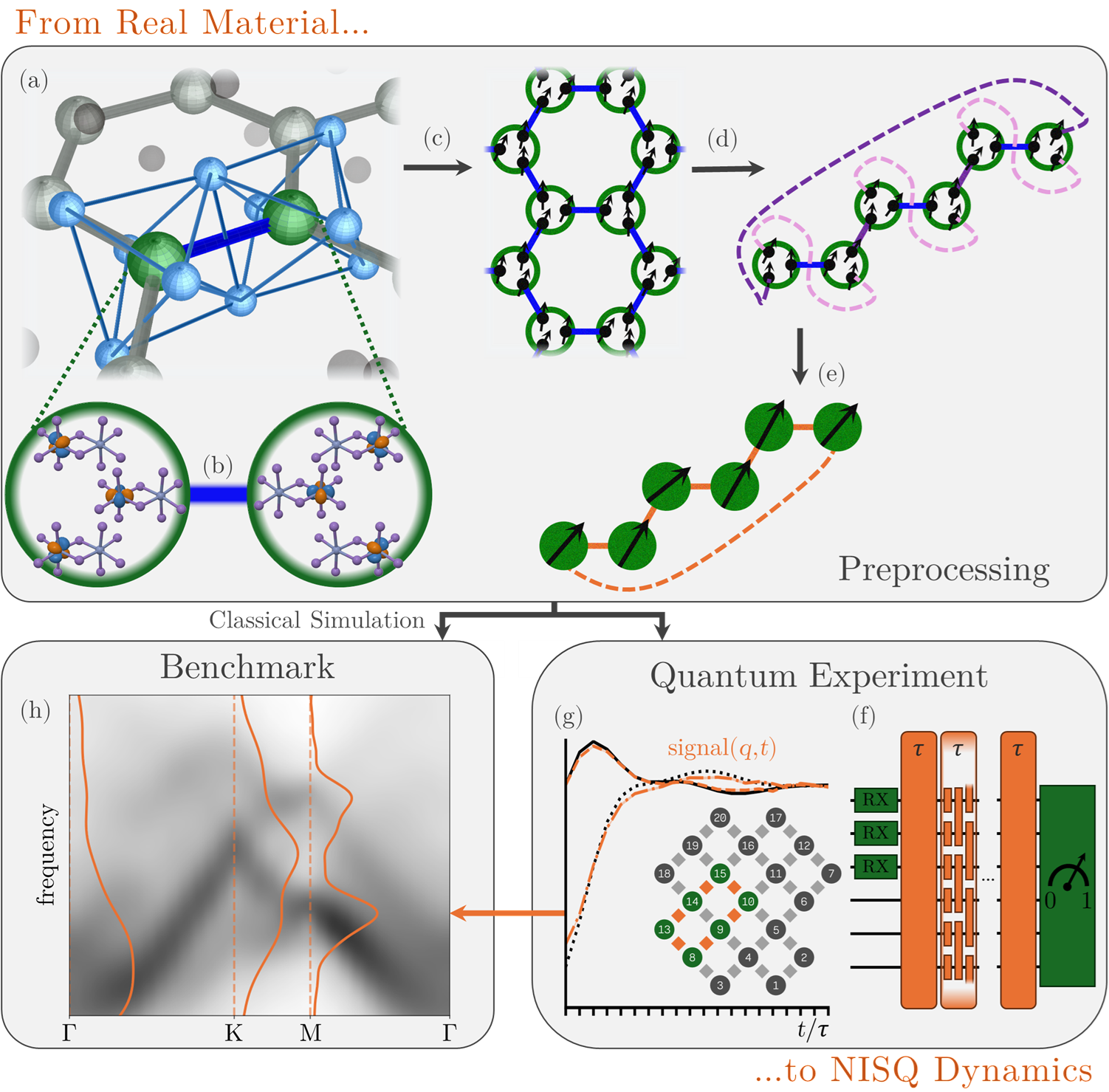}
\caption{Overview of the full workflow. We 
(a) use the structural information of a dimer of two chromium atoms (green) plus their Halide environments (light blue) inside a monolayer (grey) and 
(b) identify three spin-like electronic orbitals per chromium site. 
For the full monolayer, this 
(c) forms a spin system on a hexagonal lattice, where each site contains a cluster of three spin-1/2, which are coupled strongly among each other and weakly with the neighboring clusters. 
We (d) select a super cell, equivalent to a chain of clusters, and encode the magnon wave vector $\boldsymbol{q}$ into the interaction between chromium sites and 
(e) project to a low-energy subspace in which each cluster is described by a \emph{effective} single spin-1/2. 
By (f) building a quenched time evolution circuit and 
(g) running it on IQM Garnet or Emerald, we obtain (h) spectral data that can be compared to classical simulations.}
\label{fig:workflow_overview}
\end{figure*}

\section{From real material to NISQ}
In the following, we outline the main steps and challenges of the full workflow sketched in \cref{fig:workflow_overview},
providing a birds-eye view of how to perform a useful quantum simulation of a real material on quantum computers available today via cloud access.
Each step is complex in its own right and explained in more detail in \cref{sec:methods}.

\subsection{From ab-initio to quantum spins} \label{sec:real_mat_to_qc}
Our starting point is the structural information of a real material with the goal of obtaining an effective quantum spin model, which can be used to simulate magnon dynamics on quantum computers available today. 
To that end, we start with the lattice structure, atomic positions and unit cell of chromium tri-halide monolayers \cite{haastrup18_c2db,gjerding21_c2db}, which are periodic 2D magnets. We select a finite cluster spanned by neighboring chromium atoms together with their respective octahedral Halide environment, i.e., a finite $\mathrm{Cr}_2 X_{10}$ ($X \in \{ \mathrm{Cl}, \mathrm{Br}, \mathrm{I} \}$) cluster. This cluster, which we refer to as \emph{dimer} in the following, captures the magnetic coupling between two chromium atoms.

For each dimer, 
we choose a suitable \emph{active space} spanned by a set of single-particle orbitals
and use parity optimization \cite{schoenauer25_spin_mapper} to identify effective spin-like orbitals, 
in the following also simply referred to as ``spins''.
This procedure detects 3 spins per chromium atom, the 3 singly occupied $t_{2g}$ orbitals of the chromium ion.
By ignoring the orbitals accounting for the coupling to the remaining electronic environment, 
we obtain a strong, isotropic Heisenberg interaction between the three spins located at the same magnetic center, here a chromium atom, and a weaker Heisenberg interaction between the spins of different magnetic centers. 

Because quantum computers available today only provide a limited number of high-quality qubits, 
we need to carefully reduce the size of the spin system while retaining the relevant low-energy physics.
The final model should allow us to extract the energies of magnetic excitations by tilting (some of) the spins away from the ground state alignment and analyzing the Fourier signal of a time-dependent observable, in our case the spin magnetization in the plane perpendicular to the ground state magnetization. 
Spatial correlations of the precessional motion of the spins can be characterized by a wave vector $\boldsymbol{q}$.
We encode the wave-vector of the excitation, characterizing its momentum, in the model,
i.e., we construct a periodic Hamiltonian and inject the wave vector / momentum $\boldsymbol{q}$ by ``twisting'' the interaction of neighboring atoms depending on the bond direction (details provided in \cref{methods:wave_vector_quench}). 
This enables the simulation of large wavelengths (small momenta) 
even if $\boldsymbol{q}$ is \emph{not} commensurate with the (super) cell used in the simulation.
To simulate more complex quenches, e.g., localized, beyond single-momentum excitations, small super cells are still necessary, such that scaling up the system to more spins still provides a benefit.

An accurate time evolution of this model 
requires small time steps to account for the strong coupling between spins on the same chromium atom
and a long total simulation time to capture the magnon dynamics governed by weaker interactions between chromium atoms, 
leading to deep circuits not suited for noisy quantum computers.
We circumvent this problem through an operator-level reduction 
that preserves the relevant low-energy sector
while compressing the Hilbert space and homogenizing energy scales.
To that end, we exploit that the strong ferromagnetic interactions on a single chromium atom 
form an effective spin-$\frac{3}{2}$ degree of freedom per magnetic site, 
whose $m\in\{\frac{3}{2}, \frac{1}{2}\}$ subspace suffices to describe the low-energy spin excitations. 
In essence, we project the six-spin Hilbert space of the dimer to a 
two-\emph{effective}-spin subspace and thus arrive at a honeycomb spin-1/2 model, with the weaker Heisenberg interaction between neighboring spins as central energy scale.

Our final step to make the model amenable to contemporary quantum computers is to use super cells for which the qubit connectivity boils down to a periodic, one-dimensional chain while still representing the hexagonal lattice.
This makes it possible to map them onto a loop of qubits on the quantum chip, such that all interactions can be implemented without additional swap operations, because the connectivity of the model directly corresponds to the connectivity of the qubits on the quantum chip.

We stress that this final model, given in \cref{final_model}, is not a toy model tailored a priori to the hardware, but a model derived from first principle calculations. In fact, each interaction of the final spin model is a combination of the original Heisenberg interaction of the clusters and the wave-vector dependent twist applied to it on the re-periodized super cell of the hexagonal lattice. 

\subsection{Quantum simulation of magnon spectra}
\label{sec:qc_magnon}
We implement the time evolution of the momentum-resolved, hardware-amenable model described in the previous section on NISQ devices available via the IQM Resonance cloud interface. For each wave vector $\boldsymbol{q}$, encoded in the spin-spin interaction, and each evolution time point, we prepare an initial ferromagnetic state with all spins aligned in the same direction and seed spin dynamics by applying rotations away from the initial alignment to the first half of the spins. Since we are rotating many spins by an angle of $0.3 \pi$ ($54^\circ$), the initial quench is beyond the linear response regime. Then, we approximate the real-time evolution with a first-order Lie--Trotter product formula and extract the spin-wave spectrum from the Fourier transform of the signal
\begin{equation}
  C_{\boldsymbol{q}}(t)\;=\;-\mathrm{i}\,\frac{1}{N}\sum_{j=0}^{N-1} S^{-}_{j, \boldsymbol{q}}(0)
  S^{+}_{j, \boldsymbol{q}}(t) \, ,
\end{equation}
where $S^{\pm}_{j, \boldsymbol{q}}(t)\;=\;\hat{S}_{j, \boldsymbol{q}}^{x}(t)\,\pm\,\mathrm{i}\,\hat{S}_{j, \boldsymbol{q}}^{y}(t)\,$ denote the expectation value of the raising and lowering operators of spin $j$ at time $t$, and the subscript $\boldsymbol{q}$ denotes the wave-vector encoded in the Hamiltonian used for the time evolution, which is discussed in more detail in \cref{methods:wave_vector_quench}. 
$S^{+}(t)$ represents the rotation of the \emph{transverse} component of the spin as a simple phase and multiplying by $S^{-}(0)$ sets the initial phase to zero.

The final model, described in \cref{sec:real_mat_to_qc}, contains a constant field acting on each qubit due to the projection into the low-energy sector, which can be removed from the simulation as discussed in \cref{methods:final_hamiltonian}.
Furthermore, we combine two error mitigation techniques to obtain accurate results on actual quantum hardware. First, we mitigate coherent two-qubit gate errors using Pauli twirling (randomized compiling) \cite{wallman2016randomizedcompiling}. By averaging over randomly twirled circuit instances, coherent errors are transformed into incoherent errors. Second, we mitigate readout bias using TREX (twirled readout-error extinction) \cite{vandenberg2022trex}.
By inserting random pre-measurement bit flips and classically inverting their effect,
TREX symmetrizes readout errors, reducing bias in the estimated expectation value.

All hardware spectra are obtained from real-time dynamics generated with a first-order Lie--Trotter product formula using 19 Trotter steps, yielding 20 time steps including the initial time. The Trotter time step $\tau$ is selected using the commutator-based criterion described in \cref{methods:time-step-selection} and is kept fixed across system sizes for a given wave vector. 
Concretely, we use $\tau = 0.374~a^{-1}$ ($\Gamma$), $\tau = 0.445~a^{-1}$ (K), and $\tau = 0.72~a^{-1}$ (M), with $a$ representing the energy of the weaker Heisenberg coupling between magnetic sites.
For each evolution time, we execute $N_{\mathrm{twirl}}=16$ twirled circuit instances with 512 shots per instance yielding 8192 measurement shots per time point. As initial state, we prepare a product state obtained from the fully polarized ferromagnetic state by applying a uniform rotation by $0.3\pi$ around the $x$ axis to the first half of the spins (sites $j=1,\dots,N/2$) to trigger the spin dynamics. The high symmetry of the spin-spin interactions in the Hamiltonian results in dense compiled circuits without qubit idling. As dynamical decoupling primarily mitigates decoherence during idling periods, it is not expected to provide a benefit here.

%% file: content/Results.tex
\section{Comparing quantum and classical results}
A central requirement for turning quantum simulation into a useful computational tool is that results obtained on quantum hardware can be verified and benchmarked against a classical baseline in regimes where such a baseline exists. This emphasis on verifiability and benchmarking is highlighted both in use-case frameworks such as ITBQ (Identify, Transform, Benchmark, show Quantum advantage)~\cite{Marthaler2025ITBQ} as well as in recent frameworks for quantum applications that stress cross-checking and reproducibility as necessary conditions for utility~\cite{Babbush2025GrandChallenge}.
In the following, we therefore compare the measured momentum-resolved spin-wave spectra obtained on IQM Resonance platform to classical reference simulations and introduce a similarity metric.

As a classical baseline, we compute real-time dynamics using a Krylov-subspace time-evolution solver and post-process the resulting time traces using the same sampling grid and Fourier-analysis pipeline as for the data obtained using the quantum computer, ensuring a comparison at fixed time window and frequency resolution. The Krylov reference calculations, which essentially simulate the full Hilbert space spanned by the qubits, are feasible up to 24 spins with our current sparse-matrix implementation, where the dominant limitation is computer memory. Simulations of larger systems would be possible by going to an optimized implementation of the Hamiltonian, but this comes typically at the expense of an increase in wall time. For larger systems we employ time-dependent DMRG (td-DMRG) \cite{Schmitteckert2004_DMRG, White2004_td-DMRG} to obtain reference time traces computed on classical hardware.

\begin{figure*}[!t]
\centering
	\includegraphics[width=1\textwidth]{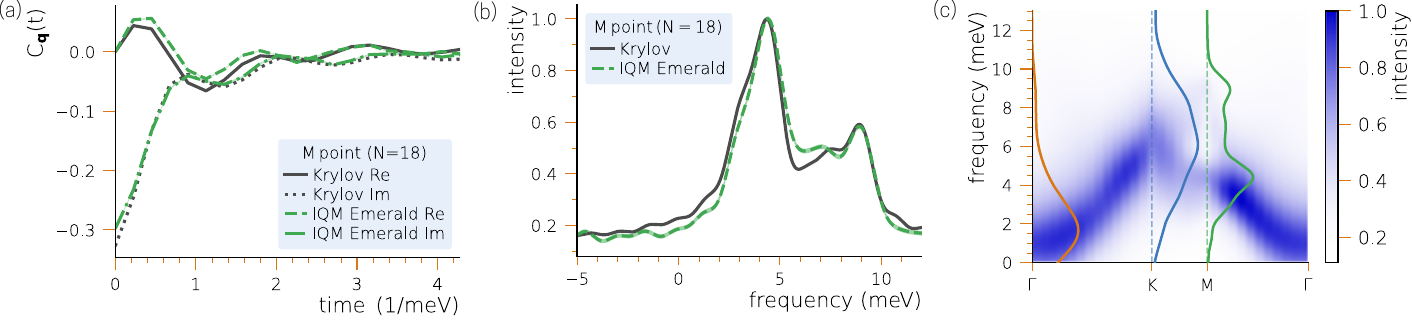}
    \caption{Benchmark of momentum-resolved spin-wave dynamics for 9 unit cells ($N=18$ spins) on IQM Emerald. \textbf{(a)} Hardware-measured time evolution of the signal $C_{\boldsymbol{q}}(t)$ at the M point obtained using the effective Hamiltonian without the local terms (cf.\ \cref{methods:final_hamiltonian}). The evolution uses 20 time points with a Trotter time step of $0.225~\mathrm{meV}^{-1}$. \textbf{(b)} Fourier signal extracted from the measured time trace; the thin, shaded band indicates statistical uncertainty obtained from bootstrap resampling, which is barely visible at this scale. Two peaks can be identified at the M point. \textbf{(c)} Fourier signals for the $\Gamma$, K, M points overlaid with a classical reference momentum-energy intensity map computed with a Krylov-subspace solver along the momentum path $\Gamma \rightarrow \mathrm{K} \rightarrow \mathrm{M} \rightarrow \Gamma$; the peaks from the quantum-hardware spectrum align with the reference spectrum.}
    \label{fig:benchmark18spins}
\end{figure*}
\cref{fig:benchmark18spins} shows benchmarks of momentum-resolved spin-wave dynamics for a supercell with 9 unit cells (18 spins) at three specific wave vectors: $\Gamma$, K, and M. At the wave vector labeled by M, the hardware time trace exhibits a clear oscillatory signal and the corresponding Fourier spectrum features two resolvable peaks with stable locations. The effect of shot noise, due to the finite number of measurements, is not visible at the scale of the plot, highlighting that incoherent errors, including also coherent errors due to our error mitigation techniques discussed in \cref{sec:qc_magnon} are the dominant error source.
The peak positions agree with the classical dispersion obtained from the Krylov solver along the path of wave vectors $\Gamma \rightarrow \mathrm{K} \rightarrow \mathrm{M} \rightarrow \Gamma$, indicating that the quantum-hardware experiment captures the dominant spectral features of the effective model.
At $\Gamma$ and K, the spectrum is dominated by a single peak, consistent with the classical reference.

\begin{figure*}[htb]
    \centering
    \vspace{\floatsep}
    \includegraphics[width=1\textwidth]{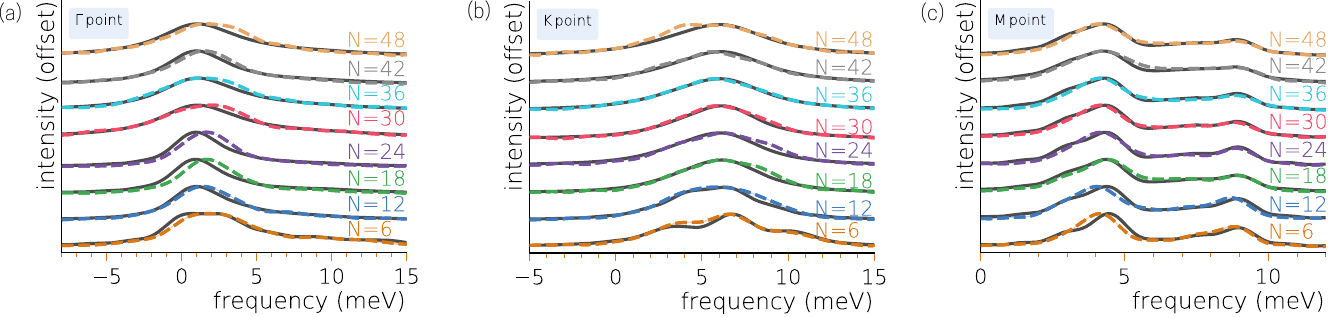}
    \caption{Scaling of the momentum-resolved spin-wave spectrum dependent on the system size.
    Fourier spectra at \textbf{(a)} the wave vector labeled by $\Gamma$, \textbf{(b)} K, and \textbf{(c)} M for super cells comprising $N=6,12,24,30,36,42,48$ spins, i.e.,  increasing super-cell sizes. Quantum-hardware spectra are obtained from real-time dynamics generated with a first-order Lie--Trotter product formula using 20 time points (19 Trotter steps); For $N\le 24$, we overlay CPU-based Krylov-subspace reference spectra computed on the same time grid as the quantum-hardware data; for $N>24$ we overlap spectra obtained using td-DMRG.}
    \label{fig:benchmarkoverlaid}
\end{figure*}
\cref{fig:benchmarkoverlaid} investigates how the spectra evolve with increasing system size at the three momenta labeled by $\Gamma$, K, and M.
We overlay the corresponding quantum-hardware spectra with data from Krylov ($N\leq24$) and DMRG ($N\geq30$) calculations, enabling a direct visual assessment of peak positions and relative intensities.
Overall, the agreement is momentum-dependent.
The M point is the most favorable case in our implementation: the circuit depth per Trotter step is smaller (depth 5 at M versus depth 6 at $\Gamma$ and K), which reduces the accumulation of gate errors and decoherence over the fixed 19 Trotter steps (20 time points including initial time). 
In contrast, the $\Gamma$ point is intrinsically challenging because the relevant modes lie close to zero frequency; resolving such low-frequency features requires longer accessible evolution times, which is difficult to realize at 
the limited circuit depth 
on current hardware. At the $\Gamma$ point, we performed additional hardware runs by increasing the number of Trotter steps and increasing the number of Pauli-twirled circuits. None of these changes improved the peak positions. We therefore attribute the peak shift primarily to hardware errors and the limited resolvability of near-zero-frequency modes at the accessible evolution times. 

\begin{figure}[tb]
\centering
	\includegraphics[width=0.9\linewidth]{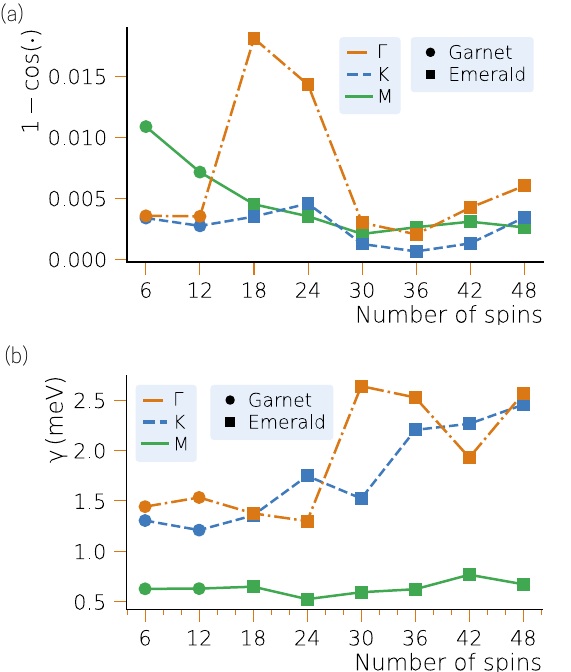}
    \caption{
    Cosine-similarity benchmark of quantum-hardware spectra against a (damped) classical reference.
    \textbf{(a)} Spectral mismatch quantified as one minus the cosine similarity, $1 - \cos(\cdot)$, between quantum-hardware and damped classical spectra versus system size. The cosine similarity is evaluated as described in \cref{methods:cosine_similarity} after interpolating both spectra to a common uniform frequency grid.
    \textbf{(b)} Extracted effective damping rate $\gamma$ obtained by multiplying the time trace obtained on classical hardware with an exponential envelope $\textrm{exp}(-\gamma t)$ and optimizing $\gamma$ to maximize the cosine similarity between the resulting ``classical'' spectrum and the corresponding quantum-hardware spectrum. No additional damping is applied to the quantum-hardware time traces.}
    \label{fig:cosine_sim}
\end{figure}
To quantify spectral agreement, we compute the cosine similarity between quantum-hardware and classical spectra as described in \cref{methods:cosine_similarity}, after interpolating both spectra to a common uniform frequency grid. 
In a nutshell, we compute the overlap of the classical and the quantum spectra. 
The quantum-hardware time traces are subject to noise and decoherence, which cause an effective decay. Because this effect is absent in the classical reference simulation, we multiply the classical time trace by an exponential envelope and optimize the corresponding effective decay rate to maximize the overlap. Hence, the fitted decay rate can be interpreted as an effective decoherence rate. 
Figure~\ref{fig:cosine_sim} summarizes both the fitted decay rates and the residual mismatch, one minus cosine similarity, as a function of system size.
The residual mismatch in Fig.~\ref{fig:cosine_sim}(a) becomes larger at the $\Gamma$ point for $N=18$ and $N=24$. This behavior agrees with the spectra in 
Fig.~\ref{fig:benchmarkoverlaid}(a), where the quantum-hardware and Krylov peak shows a relative shift. Since the $\Gamma$-point lies close to zero energy, its accurate resolution requires long evolution times and is therefore particularly sensitive to gate errors. The effective decoherence rates in Fig.~\ref{fig:cosine_sim}(b) are particularly small at the $M$ point, which is consistent with the reduced circuit depth per Trotter step. 
Additionally, it has to be kept in mind that data were collected on two distinct devices, Garnet and Emerald, over several weeks under varying device calibrations (summarized in Table~\ref{tab:iqm_calibrations}). Hence, fluctuations in hardware parameters also contribute to observed variations in the decoherence rates.

\begin{figure}[tb]
\centering
  	\includegraphics[width=0.9\linewidth]{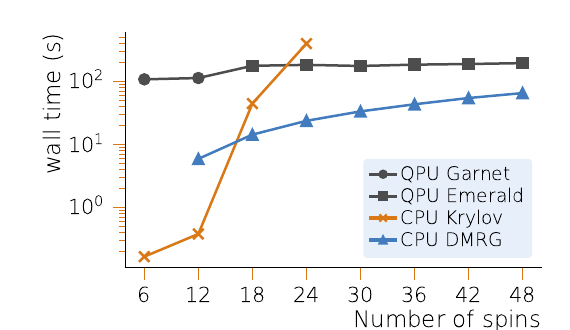}
  	    \caption{%
   Total wall time required to generate a momentum-resolved spin-wave spectrum as a function of system size. For each method, the reported runtime is averaged over the $\Gamma$, $K$, and $M$ points. Quantum-hardware runtimes are compared with classical reference calculations based on Krylov-subspace time evolution (shown up to $N=24$ spins) and DMRG (shown up to $N=48$ spins, using a discarded-entropy threshold $\delta S = 0.1$; see Appendix~\ref{methods:dmrg_calculation}). Both the Krylov and DMRG calculations are performed with the same time-evolution parameters as used in the corresponding quantum-hardware runs. Over the investigated range, the quantum-hardware runtime is approximately constant, whereas the classical runtimes increase with system size. The Krylov runtime exceeds the quantum-hardware wall time at $N=24$, while the averaged DMRG runtime remains below the quantum-hardware wall time for all system sizes shown. Since the classical simulations retain coherent oscillations over longer times, whereas the hardware signals decay earlier due to noise and decoherence, the absolute runtimes are not directly comparable. This figure should therefore be interpreted as a comparison of \emph{scaling behavior}; absolute runtimes depend on the chosen classical algorithms, their implementation and optimization, as well as hardware-specific experimental overheads.
    }
    \label{fig:scaling}
\end{figure}
Finally, we report the wall times of the quantum experiments and the classical reference simulations in \cref{fig:scaling}. The concrete values of the classical and quantum wall times depend on implementation details and, while they represent the actual simulation times for this study, should be taken with a pinch of salt. Nevertheless, two different scaling behaviors can be observed for our implementations: while the classical Krylov and DMRG computation times increase with the system size, the wall time of our quantum experiment is constant in good approximation. We point out that the DMRG wall times reported in this plot are obtained by reducing the accuracy of the DMRG calculations to shorten the wall time, while still being more accurate then the quantum computer results. The DMRG reference calculations shown in \cref{fig:benchmarkoverlaid} are obtained at a higher accuracy, to ensure converged reference data (cf.\ \cref{methods:dmrg_calculation}), resulting in a significantly increased wall time.

%% file: content/Conclusion.tex
\section{Towards quantum utility}
This work demonstrates that cloud-accessed quantum processors can produce physically meaningful results for real materials with quenches outside the linear-response regime. We showed how to setup quantum computations on actual hardware, starting from first-principles electronic structure calculations for chromium tri-halides, operating in a regime that is challenging to simulate on conventional computers. 
By simulating spin-wave spectra at various wave vectors representing high-symmetry points with near-constant quantum wall time, we validate that utility is achievable even without quantum advantage.

Challenges remain: the rather low number of 20 Trotter time steps which can be performed before the signal is damped away by noise limits the spectral resolution, in particular for near-zero frequency modes, and costs of quantum computing runs need to sink to make simulating the full dispersion more viable. Furthermore, it remains an open question whether quantum simulations on truly two-dimensional super cells, expected to be harder to simulate classically using DMRG, will be achievable in the near future. 
In the present work, swapping could be avoided due to directly mapping a 1D ring onto the connectivity of the quantum hardware, which simultaneously allowed for an efficient simulations using td-DMRG.

The derived spin-chain Hamiltonian also serves as a great model for application-driven benchmarks and future works could incorporate experimentally controllable parameters of the material, such as strain, doping or external fields. Extend this workflow to true two-dimensional effective models, meaning super cells created extending the unit cell along both lattice vectors, provides an interesting direction to explore for turning quantum utility into actual advantage. In particular, other hardware architectures such as ion traps or neutral atom devices may be better suited to reduce the number of additional SWAP operations in those two-dimensional settings.

Ultimately, this work demonstrates how far quantum computers have come: it is now possible to simulate real materials with real physical relevance, using real hardware.
This is possible simply through commercially available cloud access, rather than limited to hardware experts with direct physical access. 
Such workflows have the potential for real utility, which lies not in speedup, but in accessibility: while powerful classical methods such as DMRG often demand specialized knowledge, domain scientists can perform standard quantum time evolution algorithms via the cloud with minimal expertise.
We are confident that--as quantum devices mature further--such workflows will evolve from proof-of-concept demonstrations into indispensable tools for scientific discovery and technological innovation.

%% file: content/Methods.tex
\section{Methods} \label{sec:methods}
\subsection{Effective spin model} \label{sec:methods_effective-model}
Our goal is to obtain an effective spin model from the structural information of a real material. This means that we start with the lattice structure, atomic positions and unit cell of chromium tri-halide monolayers \cite{haastrup18_c2db,gjerding21_c2db}. A common approach is to use the magnetic force theorem \cite{Liechtenstein1987_magnetic_force_theorem} to extract the coupling constants for a \emph{classical} Heisenberg-like spin model \cite{He2021_TB2J,Esteras2022_CrSBr_TB2J} or to compute the magnon spectral function using time-dependent density functional theory (TD-DFT)\cite{Rubio2020_TDDFT_FeCoNi,Baroni2023_TDDFT_CrI3,Gorni2023_TDDFT_CrI3_magnon_phonon} or many-body perturbation theory via the solution of the Bethe-Salpeter equations (BSE) \cite{Bluegel2016_MBPT_FeCoNi,Olsen2021_MBPT_CrI3,MolinaSanchez2025_CrX3_BSE}. However, since we strive to simulate the spin model on a quantum computer, we explore a route for extracting a \emph{quantum} Heisenberg-like spin model directly from the first-principles electronic Hamiltonian. We achieve this by identifying spin-like orbitals, in the following also simply referred as ``spins'', among the single-particle orbitals of the 2D magnet, together with some orbitals accounting for the coupling of these spins to the (remaining) electronic \emph{environment}.

\subsection{Identification of spins}
Starting point is the first-principles electronic Hamiltonian
\begin{align}
    \hat{H} = \sum_{ij,\sigma} \mathrm{T}_{ij} \hat{\phi}^\dagger_{i,\sigma} \hat{\phi}^{\phantom{\dagger}}_{j,\sigma} +
    \frac{1}{2} \sum_{ijkl;\sigma\tau} \mathrm{V}_{ijkl} \hat{\phi}^\dagger_{i,\sigma} \hat{\phi}^\dagger_{j,\tau} \hat{\phi}^{\phantom{\dagger}}_{k,\tau} \hat{\phi}^{\phantom{\dagger}}_{l,\sigma} \,, \label{ab_initio_Hamiltonian}
\end{align}
with the one-body contribution
\begin{align}
    \mathrm{T}_{ij} = \int\!\!\mathrm{d}^3r \phi^\star_i(\boldsymbol{r}) \Big( -\frac{\hbar^2}{2m} \nabla^2 + v(\boldsymbol{r}) \Big) \phi^{\phantom{\star}}_j(\boldsymbol{r}) \,, \label{hopping_matrix}
\end{align}
accounting for the kinetic energy of the electrons, the interaction of electrons and nuclei, $v(\boldsymbol{r})$, and the Coulomb interaction between electrons
\begin{align}
    \mathrm{V}_{ijkl} = \frac{e^2}{4 \pi \varepsilon_0} \iint\!\!\mathrm{d}^3r\mathrm{d}^3r^\prime \frac{\phi^\star_i(\boldsymbol{r}) \phi^\star_j(\boldsymbol{r}^\prime)
    \phi^{\phantom{\star}}_k(\boldsymbol{r}^\prime) \phi^{\phantom{\star}}_l(\boldsymbol{r})}{|\boldsymbol{r} - \boldsymbol{r}^\prime |} \,. \label{interaction_matrix}
\end{align}
The matrix elements in Eqs.\ \eqref{hopping_matrix} and \eqref{interaction_matrix} are defined in terms of \emph{orthogonal} single-particle orbitals $\phi^{\phantom{\star}}_{j}(\boldsymbol{r})$.

To identify spin-like orbitals we use the concept of the so-called parity 
\begin{equation} \label{parity_definition}
    \hat{P}_j = (-1)^{\hat{n}_{j\uparrow} + \hat{n}_{j\downarrow}}\,,\,\,\,\,
    \hat{n}_{j\sigma} = \hat{\phi}^\dagger_{j \sigma} \hat{\phi}^{\phantom{\dagger}}_{j \sigma} \,,
\end{equation}
which depends on the \emph{choice} of single-particle orbital basis \cite{schoenauer25_spin_mapper}. The virtue of the parity is that its expectation value $P_j \in \left[-1, 1\right]$, where the value $-1$ is only achieved if the orbital is singly occupied in \emph{all} Slater determinants (SDs) contributing to the wave function, while it is $1$ if the orbital is empty or doubly occupied in \emph{all} SDs contributing to the wave function \cite{Shirazi2024_orbitalparities}. This has to be contrasted with the expectation value of the occupation, $\hat{n}_j = \hat{n}_{j\uparrow} + \hat{n}_{j\downarrow}$, of orbital $j$, which takes values of zero / two if the orbital  is unoccupied / doubly occupied in all SDs contributing to the wave function, but can be $1$ if \emph{either} the orbital is singly occupied in all SDs \emph{or} there is an equal number of unoccupied and doubly occupied SDs in the wave function. 
Thus, the parity $P_j$ provides the criterion of choice for finding single-occupied orbitals. Orbitals with parity (close to) $-1$ can be considered as spin-$\frac{1}{2}$ degrees of freedom, freezing out the electronic charge as they do not hop from and to other orbitals. Since the parity depends on the choice of single-particle basis, we find a basis which minimizes the parity for as many orbitals as possible and identify these orbitals as spins.

\subsection{Ab-initio calculations} 
The expectation value of the parity can be rewritten as
\begin{equation}
    \big\langle \hat{P}_j \big\rangle = \big\langle 1 - 2 \left(\hat{n}_{j \uparrow} + \hat{n}_{j \downarrow} \right) + 4 \hat{n}_{j \uparrow} \hat{n}_{j \downarrow} \big\rangle \, . \label{parity_expectation}
\end{equation}
Hence, it can be computed knowing the one- and two-body reduced density matrices (RDM) of the system. To compute the 1RDM and 2RDM we employed the so-called CASSCF method, as implemented in the PySCF code \cite{sun18_pyscf,sun20_pyscf} using Ahlrich's \emph{def2-tzvp} basis set, with effective core potentials for Iodine \cite{Weigend2005_def2}. CASSCF treats correlations within a suitable \emph{active space} spanned by a set of single-particle orbitals. The remaining orbitals are either doubly occupied or unoccupied and thereby will not host any spins. The selection of the initial active space has been proposed by the Active Space Finder (ASF) \cite{ASF2024_github}. To run the wave-function-based CASSCF calculation we extracted a finite cluster from the periodic 2D solid.
Since we are mainly interested in the magnetic coupling between two chromium atoms, we chose two neighboring chromium atoms together with their respective octahedral halide environment, i.e., a finite $\mathrm{Cr}_2 X_{10}$ ($X \in \{\mathrm{Cl}, \mathrm{Br}, \mathrm{I} \}$) cluster, which we refer to as \emph{dimer} in the following. 

We simulate a \emph{charged} dimer, with 4 extra electrons, to account for the truncated ionic bonds between the halides and the magnetic centers.
The converged active space comprises five 3d orbitals for each chromium atom and four additional p-like orbitals from the halide ligands. The spin-finding procedure detailed above detects 3 spins per chromium atom, corresponding to the three $\mathrm{t_{2g}}$ orbitals, which lower their energy due to the octahedral crystal-field splitting. The remaining orbitals of the active space are delocalized molecular orbitals comprised of the two $\mathrm{e_{g}}$ orbitals per chromium atom and p orbitals from the halides. Hence the spin identification provides us with a choice of single-orbital basis for the active space, separating it into \emph{spins} and \emph{environment}. 

The simplest effective spin model is then obtained by ignoring the environment orbitals and only considering terms in the ab-initio Hamiltonian involving the spins. This yields an isotropic Heisenberg interaction of strength $A$ between the three spins located at the same magnetic center (chromium atom) and a weaker Heisenberg interaction, $a$, between the spins of different magnetic centers. The interaction is isotropic, since, in this simple model, it is exclusively due to the \emph{bare} Coulomb interaction between the electrons. 

\subsection{Spin model analysis}
Due to the aforementioned Coulomb exchange mechanism we find a strong ferromagnetic on-site coupling $A_\mathrm{Cl} = 443.8\,\mathrm{meV}$, $A_\mathrm{Br} = 438.8 \,\mathrm{meV}$ and $A_\mathrm{I} = 429.9\,\mathrm{meV}$ for $\mathrm{Cr}_2 \mathrm{Cl}_{10}$, $\mathrm{Cr}_2 \mathrm{Br}_{10}$,  and $\mathrm{Cr}_2 \mathrm{I}_{10}$, respectively. The inter-site coupling is $\tilde{a}_\mathrm{Cl} = 3.4\,\mathrm{meV}$, $\tilde{a}_\mathrm{Br} = 3.2\,\mathrm{meV}$ and $\tilde{a}_\mathrm{I} = 3.0\,\mathrm{meV}$ for the 3 halides considered in this work. The trend across the halides for the inter-site coupling can be explained by the varying lattice constant when replacing the halides and considering that in the simplest model the interaction is exclusively due to the direct Coulomb interaction of the spin-like orbitals. We can extract a \emph{renormalized} inter-site coupling by computing total energy differences for the \emph{dimer} at different values for the total $\hat{S}^2$ operator under the assumption that the dimer can be decomposed into two magnetic centers with spin-3/2. The renormalized inter-site coupling accounts for an additional magnetic coupling due to the so-called \emph{superexchange} mechanism mediated by the halides, leading to $a_\mathrm{Cl} = 2.7 \,\mathrm{meV}$, $a_\mathrm{Br} = 3.1 \,\mathrm{meV}$ and $a_\mathrm{I} = 3.7\,\mathrm{meV}$, reversing the energetic ordering of the inter-site couplings when changing the halides. These renormalized inter-site couplings are in decent agreement with recent experimental~\cite{Chen2022_CrCl3_topological_neutron_scattering} and theoretical~\cite{MolinaSanchez2025_CrX3_BSE} results, keeping in mind that $a = \frac{9}{4} J$, with $J$ being the nearest neighbor coupling typically reported considering a (classical) spin-3/2 model. Concretely we have $J_\mathrm{Cl} = 1.21\,(0.95)\,\mathrm{meV}$, $J_\mathrm{Br} = 1.38\,(1.36)\,\mathrm{meV}$ and $J_\mathrm{I} = 1.64\,(2.01)\,\mathrm{meV}$, where the values in parenthesis are experimental results taken from Ref.\ \cite{Chen2022_CrCl3_topological_neutron_scattering}.

\subsection{Re-periodization}
To construct a periodic Hamiltonian representing the 2D magnet we ``re-periodize'' the model based on the interaction parameters, which have been extracted from the ab-initio calculation for the finite dimer. The three spins located on a single magnetic center (chromium atom) are ferromagnetically coupled by a spin-spin interaction of strength $A$. This \emph{strong} coupling expresses the fact that \emph{local} exchange favors the alignment of (electron) spins, essentially forming an effective spin-$\frac{3}{2}$ degree of freedom. The \emph{weaker} spin-spin coupling, $a$, is assigned to the nearest-neighbor bonds between magnetic centers on a hexagonal lattice. The ab-initio calculation \emph{for the dimer} does not provide us with any information about how strongly the three local spins individually are coupled to the spins of the three nearest magnetic centers, hence we \emph{choose} to couple one of the three local spins to exactly one of the nearest magnetic centers. 
The resulting Hamiltonian, comprised of 3 spin-1/2 per magnetic site, can be written as
\begin{align}
    \hat{H}_{s} = & - A \sum_{j} \Big( \hat{\boldsymbol{S}}_{j, a} \cdot \hat{\boldsymbol{S}}_{j, b}
    + \hat{\boldsymbol{S}}_{j, b} \cdot \hat{\boldsymbol{S}}_{j, c} + \hat{\boldsymbol{S}}_c \cdot \hat{\boldsymbol{S}}_{j, a} \Big) \nonumber \\
    & - a \sum_{j \in \mathcal{A}}\Big( \hat{\boldsymbol{S}}_{j, a} \cdot \hat{\boldsymbol{S}}_{j_a, a}
    + \hat{\boldsymbol{S}}_{j, b} \cdot \hat{\boldsymbol{S}}_{j_b, b} + \hat{\boldsymbol{S}}_c \cdot \hat{\boldsymbol{S}}_{j_c, c} \Big) ~, \label{H_spin}
\end{align}
where the first sum runs over all magnetic sites, while the second sum is restricted to run only over the $\mathcal{A}$ sites of the bipartite honeycomb lattice. Furthermore, the three spin-1/2 on lattice site $j$ are labeled by $a, b, c$ and the index $j_x$ ($x \in a, b, c$), denotes the three nearest neighbors of lattice site $j$, which are on the $\mathcal{B}$ sites of the bipartite lattice.

\subsection{Initial quench and wave-vector dependence} \label{methods:wave_vector_quench}
The ground state of the materials considered in this work is a state with all spins aligned, due to the ferromagnetic coupling between the spins. Our goal is to perturb the system and extract the excitation energies from the Fourier signal of a time-dependent observable. To this end we tilt (some of) the spins away from the alignment.
A typical excitation for magnetic materials are so-called magnons, which are collective excitations, where the spins precess around a common quantization axis with a fixed frequency (energy). Spatial correlation of the spins can be characterized by a wave vector, which describes how the rotation angle changes in the plane of the precessional motion when changing the position of the spin in the magnetic lattice.

The final ``twist'' of the effective model is the way we encode the wave vector of the magnetic excitation when quenching the system to trigger the dynamics. Consider the following ingredient for a \emph{spatial} Fourier transform of the spin raising operator
\begin{align}
    e^{\mathrm{i} \boldsymbol{q} \cdot \boldsymbol{R}_j} S^{+}_{j} 
    & = \Big( \cos\!\big[\boldsymbol{q} \cdot \boldsymbol{R}_j\big]
    + \mathrm{i} \sin\!\big[\boldsymbol{q} \cdot \boldsymbol{R}_j\big] \Big)
    S^{+}_{j} \nonumber \\
    & = \Big( \cos\!\big[\boldsymbol{q} \cdot \boldsymbol{R}_j\big] S^{x}_{j}
    - \sin\!\big[\boldsymbol{q} \cdot \boldsymbol{R}_j\big] S^{y}_{j}\Big) \nonumber \\
    & \phantom{=} + \mathrm{i} \Big(\sin\!\big[\boldsymbol{q} \cdot \boldsymbol{R}_j \big] S^{x}_{j}
    + \cos\!\big[\boldsymbol{q} \cdot \boldsymbol{R}_j\big] S^{y}_{j}\Big)
    \nonumber \\
    & = S^{x}_{j, \boldsymbol{q}} + \mathrm{i} S^{y}_{j, \boldsymbol{q} }
    = S^{+}_{j, \boldsymbol{q}} ~, \label{Fourier_phase_to_rotation}
\end{align}
which shows that the phase factor of the Fourier transform can be viewed as a wave-vector-dependent rotation of the spin operator around the $z$-axis. This means that instead of measuring the rotated operators $S^{x}_{j,\boldsymbol{q}}$ and $S^{y}_{j, \boldsymbol{q}}$, we can perform a ``picture change'' and measure the regular operators and instead transform the Hamiltonian, used to propagate the system, by the (inverse) of this position dependent rotation of the spins around the $z$-axis--this leads to the encoding of the wave vector in the spin-spin interaction.

This construction can be compared to Bloch's theorem, which allows for the computation of the (band-)dispersion of electrons in an infinite, periodic solid by only simulating the elementary unit cell of the material, using suitable boundary conditions accounting for the (quasi) momentum of the electron \cite{Ashcroft1976_Ashcroft_mermin}. In a nutshell, the wave vector of the magnetic excitation is encoded in how the $x$- and $y$-components of the spin operators are aligned for connected magnetic centers. To highlight the details, let us consider two spins, $\hat{\boldsymbol{S}}_i$ and $\hat{\boldsymbol{S}}_j$ coupled by an interaction $\underline{\boldsymbol{a}}$, i.e., their interaction is characterized by the coupling term $- \hat{\boldsymbol{S}}_i \cdot \underline{\boldsymbol{a}} \cdot \hat{\boldsymbol{S}}_j$ in the Hamiltonian. $\underline{\boldsymbol{a}}$ is a $3\times3$ matrix,
\begin{equation}
    \underline{\boldsymbol{a}} = \begin{pmatrix} 
        a_\perp & 0 & 0 \\
        0 & a_\perp & 0 \\
        0 & 0 & a_\parallel \\
    \end{pmatrix}\,,
\end{equation}
where we allow the interaction along the $z$-axis, $a_\parallel$, to differ from the interaction in the $x$-$y$-plane, $a_\perp$. When simulating an excitation with wave vector $\boldsymbol{q}$, we enforce that the spin on site $j$ is rotated by an angle of $\phi_{ij} = \boldsymbol{d}_{ij} \cdot \boldsymbol{q}$ around the z axis (compared to the spin on site $i$), where $\boldsymbol{d}_{ij} = \boldsymbol{r}_j - \boldsymbol{r}_i$ is the \emph{displacement} vector from site $i$ to site $j$. The \emph{wave-vector-adapted} coupling matrix then reads
\begin{equation}
    \underline{\boldsymbol{a}}_{ij} = \begin{pmatrix} 
        \phantom{+} a_\perp \cos\phi_{ij} & a_\perp \sin\phi_{ij} & 0 \\
        -a_\perp \sin\phi_{ij} & a_\perp \cos\phi_{ij} & 0 \\
        0 & 0 & a_\parallel \\
    \end{pmatrix}\,.
\end{equation}
This wave-vector-adapted spin-spin interaction, together with the tilting of the spins away from the $z$-axis, constitutes the initial quench setting the spin system in motion. It allows us to simulate arbitrary wave vectors for the magnetic excitation without the need for going to large super cells. However, when using super cells we can choose an initial tilting of the spins which \emph{is not commensurate} with the elementary cell to investigate quenches, which cannot be simulated by restricting the simulation to the elementary cell.

\subsection{From six spin-1/2 to two effective spin-3/2 per unit cell}
We study an array of $N$ unit cells, each containing two magnetic centers (three spin‑$1/2$’s per magnetic center), i.e., $6N$ spins in total. Within each magnetic center the Heisenberg exchange selects a ferromagnetic $S=3/2$ quartet as the ground multiplet, separated from the remaining states by a gap set by $A$. In our model, all other couplings are smaller by roughly three orders of magnitude, so virtual admixture of the remaining states is strongly suppressed and the quartet manifold controls the low‑energy physics.

A transformation to the coupled total-spin basis reorganizes each magnetic center from the basis of three separate spin-1/2 ($\mathcal{S}_{1/2}$), into the total-spin basis, i.e., $\mathcal{S}_{1/2} \otimes\mathcal{S}_{1/2} \otimes \mathcal{S}_{1/2} \to \mathcal{S}_{3/2} \oplus \mathcal{R} $. In this basis, we project every magnetic center onto its spin-3/2 ($\mathcal{S}_{3/2}$) quartet and ignore $\mathcal{R}$. 
Denoting by $\hat U_{\mathrm{CG}}$ the unitary change of basis defined by the Clebsch--Gordan coefficients, the resulting effective Hamiltonian is
\begin{equation}
	 \hat H^{(Q)} \; \equiv \; \hat H_{S=3/2} \;=\; \hat P_Q\,\hat U_{\mathrm{CG}}\,\hat H\,\hat U_{\mathrm{CG}}^{\dagger}\,\hat P_Q \, .
\end{equation}
The quartet‑only model has $2N$ effective spin‑$3/2$ sites (four levels per site). Crucially, the sequence respects magnetization: if the microscopic Hamiltonian conserves the total spin in z-direction, $\hat S^z_{\mathrm{tot}}$, then so does $\hat H^{(Q)}$,
\begin{equation}
	[\,\hat H^{(Q)},\,\hat S^z_{\mathrm{tot}}\,]=0,
	\qquad
	\hat S^z_{\mathrm{tot}}=\sum_{k=1}^{2N} \hat S_k^z \, .
\end{equation}
This U(1) symmetry block‑diagonalizes $\hat H^{(Q)}$ into sectors of constant $\langle \hat S^z_{\mathrm{tot}} \rangle = M$.

\subsection{Hardware‑efficient magnetization truncation}
To minimize circuit depth and gate count, we truncate the effective spin‑$3/2$ of each magnetic center to its two lowest magnetization levels $\langle \hat S^z \rangle = m\in\{\tfrac{3}{2},\tfrac{1}{2}\}$. This is a local projection $\hat P$ applied on all $2N$ sites, yielding a two‑level‑per‑site Hamiltonian
\begin{equation}
	\hat H^{(2)} \; \equiv \; \hat H_{S=3/2;\,m>0} \;=\; \hat P\,\hat H_{S=3/2}\,\hat P.
\end{equation}
In practice, this step removes amplitude only in sectors with $M$ above the one‑magnon band, so it leaves most physics intact while producing a model compatible with present‑day hardware.

Because $\hat H^{(Q)}$ commutes with $\hat S^z$, the truncated space contains the entire sector of minimal magnetization. The fully polarized state is exact and survives unchanged,i.e., the product state with all effective spins at $m=3/2$,
\begin{equation}
	\big|{\mathrm{FM}}\big\rangle \;=\; \bigotimes_{r=1}^{2N} \big|3/2,3/2\big\rangle_r \, ,
\end{equation}
is an eigenstate of $\hat H$, of $\hat H^{(Q)}$, and of $\hat H^{(2)}$.

All configurations with exactly one site at $m=1/2$ and the others at $m=3/2$ (dimension $2N$) lie completely inside the truncated space. Since $\hat H^{(Q)}$ is block‑diagonal in total magnetization, its action on this band is untouched by the projection, and $\hat H^{(2)}$ reproduces these $2N$ eigenvalues exactly.

Together, the spectra of $\hat H^{(Q)}$ and $\hat H^{(2)}$ share $1+2N$ eigenvalues (the fully polarized level plus the one‑magnon band). Higher‑magnon sectors generally involve basis states at $m=-1/2$ or $-3/2$ that are discarded; after truncation, level repulsion with those missing states is removed, so these energies may shift.

\subsection{Hamiltonian after projection} \label{methods:final_hamiltonian}
After the operator-level projection and the subsequent two-level truncation per magnetic center, the dynamics are described by a nearest-neighbor spin-$1/2$ model on a one-dimensional ring of $N$ effective spins. We label sites by $i\in\{0,\dots,N-1\}$ and impose periodic boundary conditions via
\begin{equation}
    i+1 \equiv (i+1)\bmod N \, .
\end{equation}
We write the Hamiltonian as a sum of spin-spin interactions,
\begin{align}
    \hat H^{(2)}
    = & \sum_{i=0}^{N-1} \Big[
        {} J_{\perp}^{(i)}
        \left(\hat S_i^x \hat S_{i+1}^x
        + \hat S_i^y \hat S_{i+1}^y\right)
        + J_{z}^{(i)} \hat S_i^z \hat S_{i+1}^z \nonumber \\
      & \phantom{\sum_{i=0}^{N-1} \Big[} + J_{\times}^{(i)}
        \left(\hat S_i^x \hat S_{i+1}^y
        - \hat S_i^y \hat S_{i+1}^x \right) \Big]
        + \hat H_\mathrm{loc} \, , \label{final_model}
\end{align}
where the coupling constants $J_\perp$ and $J_\times$ depend on the wave-vector encoded in the interactions. Moreover, the Hamiltonian contains local terms arising from the projection,
\begin{equation}
    \hat H_{\mathrm{loc}}=\sum_{i=0}^{N-1} h_i \hat S_i^z .
\end{equation}
To reduce the relevant energy scale entering the Trotterization, we subtract the uniform (spatially averaged) component of the local terms,
\begin{equation}
    \alpha = \frac{1}{N}\sum_{i=0}^{N-1} h_i,
    \qquad
    \hat H' = \hat H^{(2)} - \alpha \sum_{i=0}^{N-1} \hat S_i^z .
\end{equation}
Equivalently, this replaces $h_i$ by the centered fields $h_i-\alpha$ while leaving all interaction terms unchanged. The model conserves total magnetization along the $z$-axis, such that the subtracted uniform term generates only a precession about the $z$-axis (i.e., a uniform frequency shift in the spectra), which can be accounted for in post-processing. Since our model is uniform, the subtraction removes it completely.

\subsection{Trotterization parameters: time-step selection}
\label{methods:time-step-selection}
To simulate real-time dynamics on hardware we use a first-order Lie--Trotter product formula. We fix the number of Trotter steps to $19$, yielding 20 sampled time points including the initial time, in order to remain within the coherence budget of current hardware. The Trotter time step $\tau$ is chosen empirically. As a starting point, we use the commutator-based estimate
\begin{equation}
    \varepsilon_{\mathrm{LT}}(\tau)
    =
    \frac{T_{\max}\tau}{2}
    \sum_{\ell<m}\bigl\|[\hat H_\ell,\hat H_m]\bigr\| ,
\end{equation}
for the centered Hamiltonian $\hat H'=\sum_{\ell=1}^L \hat H_\ell$, with $T_{\max}=n_T\tau$. For the smallest system considered ($N=6$ spins), we choose $\tau$ such that this estimate is approximately $0.7$. We then validate and refine this choice by comparing the spectra obtained from ideal (noise-free) Lie--Trotter circuit simulations to the corresponding Krylov-subspace reference calculations over the experimentally relevant time window. In practice, the chosen time step provides a good compromise between limiting Trotter errors and reaching sufficiently long evolution times for spectral resolution. We further verified that modest increases or decreases of $\tau$ lead to less favorable agreement with the Krylov reference in the extracted spectra, and that the same qualitative dependence on $\tau$ is also reflected in the hardware data. 

\subsection{Cosine similarity between quantum and classical spectra}
\label{methods:cosine_similarity}
To quantify spectral agreement between quantum-hardware data and classical reference calculations we use the cosine similarity~\cite{Fratus2025NMRQCvsClassical} between two (real-valued) spectral functions $C_a(\omega)$ and $C_b(\omega)$,
\begin{equation}
    \cos\theta_{ab}
    =
    \frac{
        \int_{-\infty}^{+\infty} C_a(\omega)\,C_b(\omega)\,d\omega
    }{
        \sqrt{\int_{-\infty}^{+\infty} C_a(\omega)^2\,d\omega}\;
        \sqrt{\int_{-\infty}^{+\infty} C_b(\omega)^2\,d\omega}
    }.
\end{equation}
In practice, we evaluate the integrals numerically on a uniform frequency grid $\{\omega_i\}_{i=1}^{N_\omega}$ spanning a range that covers the support of both spectra. Approximating each integral by a Riemann sum yields
\begin{equation}
    \cos\theta_{ab}
    \approx
    \frac{
        \sum_{i=1}^{N_\omega} C_a(\omega_i)\,C_b(\omega_i)
    }{
        \sqrt{\sum_{i=1}^{N_\omega} C_a(\omega_i)^2}\;
        \sqrt{\sum_{i=1}^{N_\omega} C_b(\omega_i)^2}
    }.
\end{equation}

For the comparisons in Fig.~\ref{fig:cosine_sim}, we compute $\cos\theta_{ab}$ between the quantum-hardware spectrum and a damped Krylov or DMRG reference spectrum. The damping is introduced at the level of the classical time trace by multiplying it with an exponential envelope prior to Fourier transformation,
\begin{equation}
    C_{\mathrm{ref}}(t)\ \mapsto\ e^{-\gamma t}\,C_{\mathrm{ref}}(t),
\end{equation}
and the decay rate $\gamma$ is chosen to maximize the cosine similarity. The quantum-hardware time traces are left unmodified. The reported mismatch is $1-\cos\theta_{ab}$.

\subsection{DMRG calculation parameters}
\label{methods:dmrg_calculation}
For comparison with systems too large to be treated in the complete Hilbert space we resort to a mature DMRG code being used for production runs for over two decades. We stress that this DMRG code has been developed to obtain accurate results for hard quantum problems and \emph{is not tuned} for fast executions to outperform quantum computers. In our work we use it to obtain reference results to evaluate the accuracy of the QC results, including the extraction of the effective damping presented in \cref{fig:cosine_sim}.
Calculations are performed by connecting the two blocks via two sites inserted between them, using the time evolution algorithm described in Ref.~\cite{Schmitteckert2004_DMRG}, which is based on a Krylov subspace expansion and does not employ a Trotter decomposition. In addition we are not relying on adaptive td-DMRG \cite{White2004_td-DMRG} but perform the time evolution at every DMRG step, reducing the run-away error in td-DMRG simulations \cite{Schmitteckert2008_correlated_transport,Schmitteckert2012_boson_dmrg}.

We start the simulation by directly initializing the tensor-product state as an exact initial state and perform a first time step. We then iteratively restart the DMRG performing 2, 3, 5, 7, 15 and finally all 19 time steps. In the benchmark calculation we increased the number of states kept to ensure that the discarded entropy $\delta S$ is below $10^{-4}$. For up to 36 sites we have checked that the results coincide with calculations with $\delta S$ below $10^{-7}$. We also observed that the results do not change significantly when lowering $\delta S$ to $0.01$, while $\delta S = 0.1$ does lead to deviations of the order of a few percent in the measured observables.

\subsection{Hardware calibration parameters}
See Table~\ref{tab:iqm_calibrations}.

\begin{table*}
\centering
\begin{tabular}{llllllllllll}
\toprule
$N$ & $\mathbf{q}$ & QPU & $\bar{\epsilon}_{1\mathrm{Q}}$ & $\bar{t}_{1\mathrm{Q}}$ [ns] & $\bar{\epsilon}_{2\mathrm{Q}}$ & $\bar{t}_{2\mathrm{Q}}$ [ns] & $\bar{\epsilon}_{\mathrm{RO}}$ & $\bar{t}_{\mathrm{RO}}$ [ns] & $\overline{T_1}$ [$\mu$s] & $\overline{T_2}$ [$\mu$s] & \text{Date}\\ 
\midrule
6 & $\Gamma$ & Garnet  & 0.88e-3 & 16 & 5.1e-3 & 40 & 2.3e-2 & 123 & 33.9 & 7.87 & 22.01.2026 \\
12 & $\Gamma$ & Garnet  & 1.3e-3 & 16 & 4.3e-3 & 40 & 2.7e-2 & 120 & 32.8 & 8.04 & 22.01.2026 \\
18 & $\Gamma$ & Emerald  & 0.93e-3 & 33.6 & 5.7e-3 & 60 & 1.6e-2 & 360 & 49.8 & 18.6 & 31.01.2026 \\
24 & $\Gamma$ & Emerald  & 0.72e-3 & 33.2 & 5.e-3 & 60 & 1.7e-2 & 360 & 48 & 15.4 & 02.02.2026 \\
30 & $\Gamma$ & Emerald  & 0.45e-3 & 24 & 5.e-3 & 60 & 1.5e-2 & 360 & 44.4 & 19.5 & 08.04.2026 \\
36 & $\Gamma$ & Emerald  & 0.53e-3 & 24 & 8.4e-3 & 60 & 1.6e-2 & 360 & 45.5 & 18.3 & 08.04.2026\\
42 & $\Gamma$ & Emerald  & 0.61e-3 & 24 & 8.5e-3 & 60 & 1.6e-2 & 360 & 44.8 & 17.6 & 08.04.2026\\
48 & $\Gamma$ & Emerald  & 1.2e-3 & 24 & 9.1e-3 & 60 & 1.8e-2 & 360 & 44.5 & 17.3 & 08.04.2026\\
6 & K & Garnet  & 0.88e-3 & 16 & 5.1e-3 & 40 & 2.3e-2 & 123 & 33.9 & 7.87 & 22.01.2026\\
12 & K & Garnet  & 1.3e-3 & 16 & 4.3e-3 & 40 & 2.7e-2 & 120 & 32.8 & 8.04 & 22.01.2026\\
18 & K & Emerald  & 0.64e-3 & 32 & 4.7e-3 & 60 & 1.7e-2 & 360 & 49.3 & 15.6 & 26.01.2026\\
24 & K & Emerald  & 0.73e-3 & 33.2 & 5.4e-3 & 60 & 1.7e-2 & 360 & 49.1 & 16.3 &  02.02.2026\\
30 & K & Emerald  & 0.45e-3 & 24 & 5.e-3 & 60 & 1.5e-2 & 360 & 44.4 & 19.5 & 08.04.2026\\
36 & K & Emerald  & 0.53e-3 & 24 & 8.4e-3 & 60 & 1.6e-2 & 360 & 45.5 & 18.3 & 08.04.2026\\
42 & K & Emerald  & 0.61e-3 & 24 & 8.5e-3 & 60 & 1.6e-2 & 360 & 44.8 & 17.6 & 08.04.2026\\
48 & K & Emerald  & 1.2e-3 & 24 & 9.1e-3 & 60 & 1.8e-2 & 360 & 44.5 & 17.3 & 08.04.2026\\
6 & M & Garnet &  1.2e-3 & 16 & 5.3e-3 & 40 & 2.3e-2 & 126 & 29.6 & 9.35 & 21.01.2026\\
12 & M & Garnet  & 1.2e-3 & 16 & 4.2e-3 & 40 & 2.7e-2 & 120 & 32.8 & 8.04 & 22.01.2026\\
18 & M & Emerald  & 0.71e-3 & 32 & 3.7e-3 & 60 & 1.7e-2 & 360 & 46.7 & 16.1 & 26.01.2026\\
24 & M & Emerald  & 0.62e-3 & 33.1 & 5.8e-3 & 60 & 1.8e-2 & 360 & 51.2 & 16.3 &  02.02.2026\\
30 & M & Emerald  & 0.75e-3 & 33.9 & 6.1e-3 & 60 & 2.1e-2 & 360 & 51.8 & 15 & 02.02.2026\\
36 & M & Emerald  & 1.1e-3 & 33.4 & 8.3e-3 & 60 & 2.2e-2 & 360 & 46.6 & 15.2 & 05.02.2026\\
42 & M & Emerald  & 1.1e-3 & 33.2 & 6.4e-3 & 59.6 & 2.2e-2 & 360 & 49.1 & 16 & 05.02.2026\\
48 & M & Emerald  & 0.81e-3 & 33.1 & 6.2e-3 & 59.7 & 2.3e-2 & 360 & 53.3 & 16 & 09.02.2026\\
\bottomrule
\end{tabular}
\caption{\textbf{Average hardware-calibration parameters for all quantum-hardware runs.}
Each row corresponds to a single run identified by the system size $N$ (number of spins in the simulated effective model) and the momentum point $\mathbf{q}\in\{\Gamma,K,M\}$ The column \emph{QPU} indicates the quantum processor used.
Reported are the average single-qubit gate error $\bar{\epsilon}_{1\mathrm{Q}}$ and gate time $\bar{t}_{1\mathrm{Q}}$ (ns), the average two-qubit \texttt{cz} error $\bar{\epsilon}_{2\mathrm{Q}}$ and gate time $\bar{t}_{2\mathrm{Q}}$ (ns), the average readout (measurement) error $\bar{\epsilon}_{\mathrm{RO}}$ and measurement time $\bar{t}_{\mathrm{RO}}$ (ns), and the average coherence times $\overline{T_1}$ and $\overline{T_2}$ ($\mu$s) over the qubits used in the corresponding run. Gate and readout averages are computed over the calibrated qubits present in the run (weighted by circuit usage), while $\overline{T_1}$ and $\overline{T_2}$ denote arithmetic means over the involved qubits. The last two column show the number of single and two-qubit gate per Trotter circuit. The two-qubit circuit depth at the $\Gamma$ and K points is 6, and at the M point it is 5. The corresponding total circuit depths are 13 at $\Gamma$ and K, and 11 at M. The number of total two-qubit gates per Trotter circuit is $3\cdot N$ at $\Gamma$, K and $(5/2)\cdot N$ at M.}
\label{tab:iqm_calibrations}
\end{table*}

%% file: adds/acknowledgements.tex
The underlying research is part of the project \emph{Algorithms for quantum computer development in hardware-software codesign} (ALQU), \url{https://qci.dlr.de/en/alqu},
which were made possible by the \emph{DLR Quantum Computing Initiative} (QCI) and the \emph{German Federal Ministry of Research, Technology and Space} (BMFTR). 

%% file: adds/data_availability.tex
Hamiltonian matrices, raw bitstring results, and derived results are available from the corresponding author upon reasonable request. Requests should include a brief description of intended use. Access will be granted within 3 months of request, subject to approval by the German Aerospace Center (DLR).

%% file: adds/code_availability.tex
Pre- and post-processing workflow scripts are available under same terms as the data.

Quantum chemistry calculations used the open-source packages PySCF \cite{sun17_pyscf} (2.10.0) and Active Space Finder \cite{ASF2024_github} (2.0.1). The DMRG benchmarks were performed using the open-source package available at \texttt{svn.tuxfamily.org/svnroot/prg/nrg}. These tools are freely available under their respective open-source licenses.

The HQS Spin Mapper (1.0.3) used in preprocessing and the HQS Quantum Solver (1.6.0) used for the classical Krylov benchmark are proprietary software; Licenses may be purchased from HQS Quantum Simulations GmbH at \texttt{cloud.quantumsimulations.de}.

%% file: adds/author_contributions.tex
F.G.E. conceived this project with input from P.Sch., M.M., and P.K.S. P.St. performed the quantum computer runs and circuit simulations. P.Sch. computed the DMRG reference data. F.G.E. developed the theoretical model and carried out the ab-initio simulations with input from B.M.S. and P.Sch. P.St., F.G.E., and G.S. wrote the manuscript with input from all authors. S.Z. supervised the project. All authors discussed the results and approved the final version of the manuscript.

%% file: adds/competing_interests.tex
The authors declare no competing interests. 